\documentstyle[prl,aps,twocolumn,epsf]{revtex}\begin{document}\iffalse  

\draft

\title{	Naturalness Bounds on Dipole Moments from New Physics }

\author{       Keiichi Akama,$^1$ Takashi Hattori$^2$ and Kazuo Katsuura$^3$}
\address{       $^{1,3}$ Department of Physics, Saitama Medical College,
                Moroyama, Saitama, Japan}
\address{       $^2$ Department of Physics, Kanagawa Dental College,
                Yokosuka, Kanagawa, Japan}
\maketitle
\begin{abstract}
Assuming naturalness that the quantum corrections to the mass 
	should not exceed the order of the observed mass,
	we derive and apply model-independent bounds on
	the anomalous magnetic moments and electric dipole moments 
	of leptons and quarks due to new physics.
\end{abstract}


\else
\draft
\twocolumn[

\widetext
\hfill SMC-PHYS-165

\hfill hep-ph/0111238

\centerline{\large\bf	Naturalness Bounds on Dipole Moments from New Physics }
\vskip 10pt
\centerline{         Keiichi Akama,$^1$ Takashi Hattori$^2$ and Kazuo Katsuura$^3$}
\centerline{\small\sl      $^{1,3}$ Department of Physics, Saitama Medical College,
                Moroyama, Saitama, Japan}
\centerline{\small\sl     $^2$  Department of Physics, Kanagawa Dental College,
                Yokosuka, Kanagawa, Japan}
\vskip 5pt

\leftskip 15mm\rightskip 15mm
\begin{abstract}
\baselineskip  = 10pt
Assuming naturalness that the quantum corrections to the mass 
	should not exceed the order of the observed mass,
	we derive and apply model-independent bounds on 
	the anomalous magnetic moments and electric dipole moments 
	of leptons and quarks due to new physics.
\end{abstract}


]

\narrowtext
\fi
In spite of its splendor of the phenomenological successes, 
	the standard model of the elementary particles
	still leaves unanswered  many fundamental questions, such as 
	the origin of the quark-lepton generations, 
	the curious pattern of their mass spectrum,
	and the unnatural fine tuning in the Higgs mass renormalization
	\cite{naturalness}.
People expect that some new physics at some near-by high energy scale, such as 
	compositeness \cite{comp}, broken supersymmetry \cite{susy}, 
extra dimensions \cite{KK}, or brane worlds \cite{brane},
	would open ways to resolve these problems.
At its early stage, signatures of the new physics might reveal themselves
	through effective non-renormalizable interactions such as 
	anomalous magnetic moments \cite{ammcomp}--\cite{ammKK} 
	and electric dipole moments \cite{edm}. 
The quantum corrections to the masses due to these effects, 
	diverge badly with an effective momentum cut-off 
	at the new physics scale.
On the other hand, masses of the quarks, leptons, 
	gauge bosons, and Higgs scalar are observed to be small or very small
	in comparison with the expected new-physics scale.
It is unnatural that the large quantum corrections accidentally cancel 
	its large bare mass to give the small or very small observed masses, 
	unless it is protected by some dynamical mechanism
	which does not work at the tree level.
This last exception is very unlikely.
Thus we can assume that the quantum contribution $\delta m_{\rm new}$ 
	from the new physics
	should not exceed the order of the observed mass $m_{\rm obs}$. 
\begin{eqnarray}
	|\delta m_{\rm new}| \le O(m_{\rm obs})  
\ \ \ (\rm naturalness \ bound) \label{nb}
\end{eqnarray}
The $\delta m_{\rm new}$ in the left hand side of (\ref{nb}) 
	is written in terms of 
	the new-physics parameters (the effective coupling constants, 
	the cutoff scales, the heavy state masses, etc.)
	and other known quantities,
	and consequently it imposes a bound on the new-physics parameters.
In fact, a relation of the type (\ref{nb}) for Higgs scalar mass
	is used to advocate the necessity of some new physics \cite{naturalness}.
An argument with (\ref{nb}) for excited states
	in the composite model was given in Ref. \cite{resonanceNB} sometime ago.
In Ref.\ \cite{AK}, two of the present authors (K.~A.\ and K.~K.)
	considered a model where the naturalness bound with (\ref{nb}) 
	is saturated solely
	by the effects of anomalous magnetic moment from some new physics.
In this paper, we apply the naturalness bound (\ref{nb}) to effective magnetic 
 	and electric dipole moments of fermions, 
	which are expected in many of the new-physics candidates 
	\cite{ammcomp}--\cite{edm},
	and we derive many useful phenomenological bounds.

\begin{figure}
\hskip2.5cm\epsfxsize=3cm\epsffile{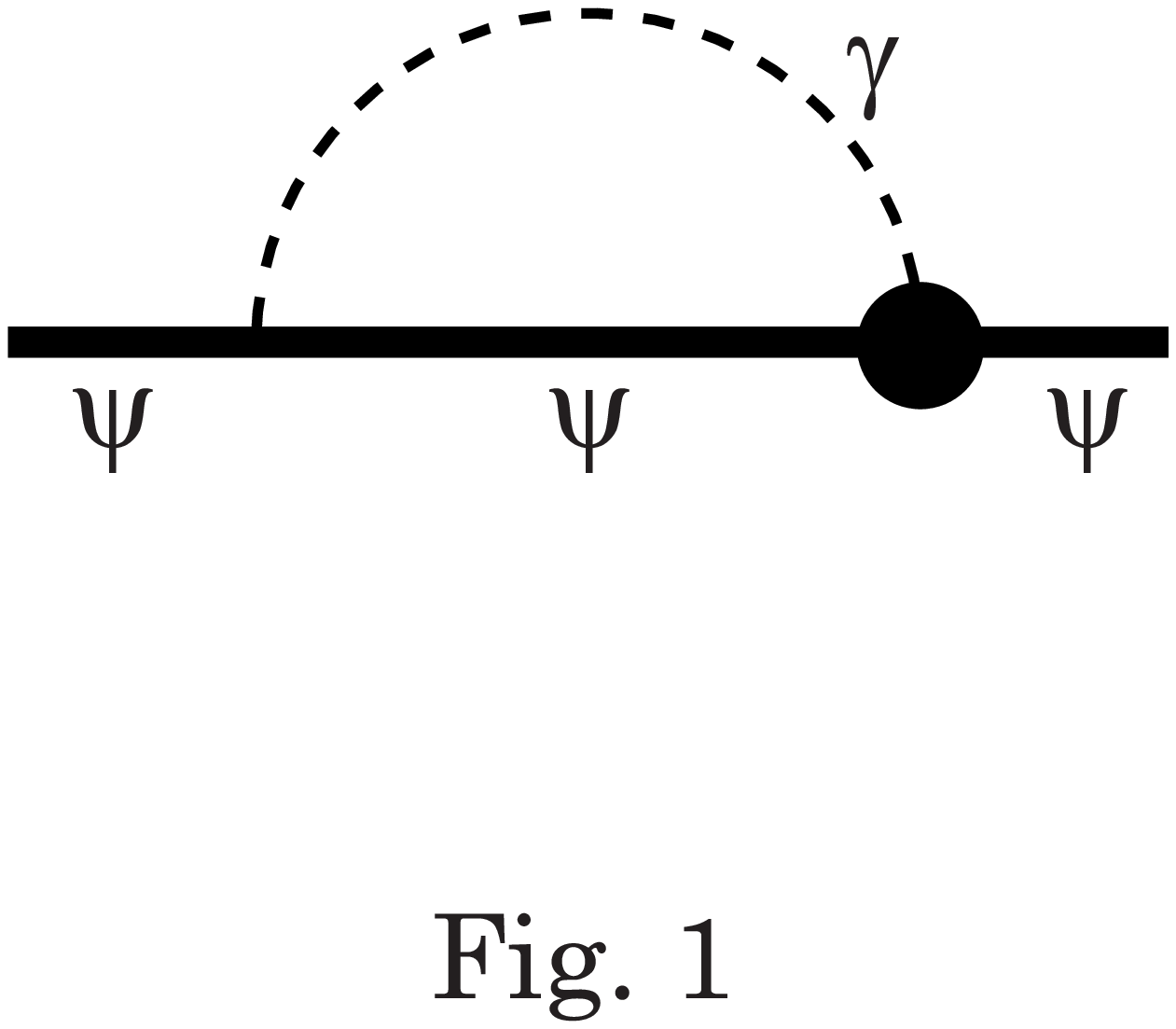}
\caption{ 
Self energy diagram for charged fermion with an anomalous vertex part, 
which is indicated by the blob. 
}
\label{f1}
\vskip 1cm
\hskip2.5cm \epsfxsize=3cm\epsffile{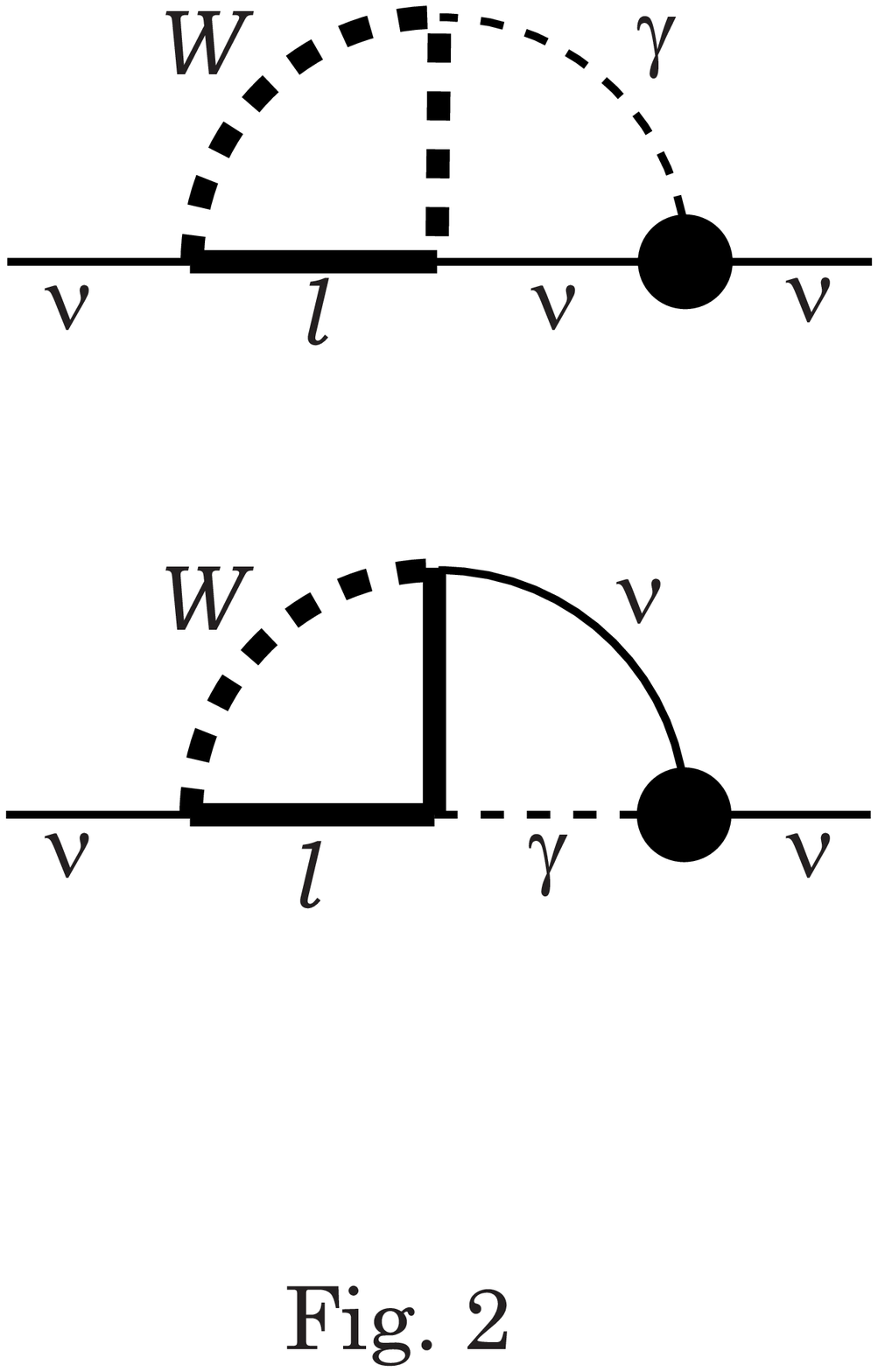}
\caption{
Self energy diagrams for neutrino with an anomalous vertex part, 
which is indicated by the blob.
}
\label{f2}
\end{figure}

We suppose that the new-physics 
 	induces the anomalous magnetic moment 
	$\mu$ and/or electric dipole moments $d$ of quark or lepton $\psi$ 
	at low energies in comparison with new-physics scale $\Lambda$. 
The latter violates CP invariance.
The effective Lagrangian for the interaction is given by
\begin{eqnarray}
	{\cal L}=-\frac{1}{2}\mu\overline\psi \sigma_{\mu\nu}F^{\mu\nu}\psi
	-\frac{1}{2}i d\overline\psi \sigma_{\mu\nu}\gamma_5 F^{\mu\nu}\psi,
\label{Lag}
\end{eqnarray}
	where $F^{\mu\nu}$ is the field strength of photon $A^\mu$.  
Though (\ref{Lag}) is a low-energy approximation for the real physics,
	we need to take into account its quantum effects 
	up to its characteristic scale.
If (\ref{Lag}) were a fundamental interaction, 
	the diagram in Fig.\,\ref{f1} would give rise to 
	a quadratically divergent contribution to the fermion mass,
	which severely violates renormalizability.
Now we argue that the internal line momenta of the diagram 
	are, in many cases, 
	effectively cut off at the characteristic scale of the new physics. 
For example, in the composite models, or the brane world models, 
	the interaction for the momenta much higher than 
	the inverse size of the composite particle extension 
	or the brane world width can no longer be expressed in the form (\ref{Lag}). 
Even though the effects of the high momenta
	should be taken into account by some other way,
	it is not through (\ref{Lag}).
Thus we cutoff the momenta as far as (\ref{Lag}) is concerned.
In the supersymmetric models,
	no real quadratic divergece in the diagram in Fig.\,\ref{f1} exists
	because they are canceled by those from the diagrams
	with the super partner internal lines. 
The symmetry, however, is broken, 
	and the contributions of the order of the breaking scale
	do not cancel,
	while those much higher than the scale cancel.
Thus the momenta are cut off at the breaking scale.

As an approximation to these existing mechanism of the momentm cutoff, 
	we insert the cutoff function
\begin{eqnarray}
	(1-\Lambda^2/q^2)^{-2}
\label{cutoff}
\end{eqnarray}
	at the photon propagator, where $\Lambda$ is the new physics scale,
	and $q$ is the photon momentum.
The approximation with (\ref{cutoff}) is sufficient for our purpose, 
	because we are concerned with the order-of-magnitude relation (\ref{nb}).
If one wants, the cutoff (\ref{cutoff}) can be done in a gauge covariant way,
	by introducing the covariant derivative regularization
	to the photon kinetic terms.
Then, it is straightforward to see that,
	due to the quantum effects via Fig.\,\ref{f1},
	the fermion mass term
	acquires the correction 
	$\overline\psi(\delta m+i\delta m_5 \gamma_5)\psi $ with
\begin{eqnarray}
	\delta m = - 3eQ\mu\Lambda^2/8\pi^2,\ \ \ 
	\delta m_5 = - 3eQd\Lambda^2/8\pi^2,
\label{deltam}
\end{eqnarray}
	where $e$ is the electromagnetic coupling constant, 
	$Q$ is the electric charge of the fermion $\psi$,
	and we have neglected small contributions compared with $\Lambda^2$.
The bare Lagrangian in general should include a $\overline\psi\gamma_5\psi$-term:
\begin{eqnarray}
	L= \overline\psi(m_0+i m_5 \gamma_5)\psi, 
\end{eqnarray}
	where $ m_0$ and $ m_{5}$ are the bare mass parameters.
The physical mass $m$ is defined as the coefficient of the $\overline\psi\psi$-term
	in the effective Lagrangian in the chirally transformed frame 
	where the $\overline\psi\gamma_5\psi$-term vanish.
Then we have 
\begin{eqnarray}
	\ m = \sqrt{(m_0+\delta m+\cdots)^2+
	(m_{5}+\delta m_5+\cdots)^2}.
\end{eqnarray}
where ``$\cdots$" indicates the other quantum corrections.
The naturalness implies $\delta m,\ \delta m_5<O(m)$, so that 
\begin{eqnarray}
	3e|Q\mu |\Lambda^2/8\pi^2<O(m),\ \ \ 
	3e|Qd|\Lambda^2/8\pi^2<O(m).
\label{nb2}
\end{eqnarray}

The relations in (\ref{nb2}) have three interesting ways of 
	phenomenological applications.
\\(i) We know from many existing experiments that 
	the new physics scale is, roughly, at least greater than 
	$\Lambda_{\rm min}$=1TeV \cite{PDG}.
Then we have the model-independent upper bounds 
\begin{eqnarray}
	|\mu|,|d|<O\left ({8\pi^2m/3e|Q|\Lambda_{\rm min}^2}\right) .\ \ \ 
\label{mu,d<}
\end{eqnarray}
If we have experimental value greater than the bound (\ref{mu,d<}),
	we would face with a serious fine tuning problem.
They often render the most stringent phenomenological bounds for $\mu$ and $d$.
\\(ii) If we know the experimental upper bound $|\mu|_{\rm max}$ or $|d|_{\rm max}$
	for $|\mu|$ or $|d|$, we have
\begin{eqnarray}
	|\kappa|\Lambda <
	O\left (\sqrt{8\pi^2m|\kappa|_{\rm max}/3e|Q|}\right).\ \ \
	(\kappa=\mu {\rm\ or\ } d) 
\label{kL<}
\end{eqnarray}
The quantity $|\kappa|\Lambda$ serves as 
	the dimensionless coupling constant 
	in perturbation expansion with the interaction Lagrangian (\ref{Lag}),
	and its smallness is desired.
\\(iii) If we have real evidences that the dipole moment $\mu$ or $d$
	deviates from the standard model predictions,
	and know the experimental lower bound $|\mu|_{\rm min}$ or $|d|_{\rm min}$
	for $|\mu|$ or $|d|$, 
	then the naturalness sets the model-independent {\it upper bound} 
	for the responsible new-physics scale $\Lambda$:
\begin{eqnarray}
	\Lambda <
	O\left (\sqrt{8\pi^2m/3e|Q||\kappa|_{\rm min}}\right).\ \ \ 
	(\kappa=\mu {\rm\ or\ } d) 
\label{L<}
\end{eqnarray}

Now we apply the bounds (\ref{mu,d<})--(\ref{L<}) 
	to the individual cases of the leptons and quarks.
We indicate the quantities for each fermion by subscripts like $\mu_e$, $d_\mu$ etc. 
Following conventions in the literature, we use
	$\delta a =\mu /(eQ/2m)$ instead of $\mu$ itself
	for the anomalous magnetic moment of charged leptons.

\noindent{\bf Muon}:  
The bound (\ref{mu,d<}) with $\Lambda_{\rm min}$=1TeV implies  
\begin{eqnarray}
	|\delta a_\mu| < O(6\times10^{-6}),\ \ 
	|d_\mu| < O (6\times10^{-19}e\rm cm), 
\end{eqnarray}
where the former is much less stringent than the experimental deviation 
	from the standard-model expectation
	recently reported by MUON $(g-2)$ collaboration \cite{mu_mu}:
\begin{eqnarray}
	\delta a_\mu = (43\pm16)\times10^{-10},    \label{exp_a_mu}
\end{eqnarray}
and the latter is comparable with the experimental bound \cite{d_mu}
\begin{eqnarray}
	d_\mu = (3.7\pm3.4) \times10^{-19}e\rm cm.    \label{exp_d_mu}
\end{eqnarray}

Then we use the bound (\ref{kL<}) with the 95\%CL(confidence level) upper bounds 
	from (\ref{exp_a_mu}) and (\ref{exp_d_mu}) to get 
\begin{eqnarray}
	|\mu_\mu|\Lambda  <O(0.0003), \ \ \ 
	|d_\mu|\Lambda  <O(0.012), \ \ \ 
\end{eqnarray}
which justifies the pertubation expansions in $\mu_\mu$ and $d_\mu$.

What is remarkable with the experimental result (\ref{exp_a_mu}) for $\delta a_\mu$  
	is that it deviates from the standard-model prediction 
	by 2.6 standard deviations,
	providing a possible signature for some new physics.
This renders us a presently only chance to use the bound (\ref{L<}), 
which sets the model-independent upper bound 
\begin{eqnarray}
	\Lambda  <O(70\rm TeV) \ \ (95\%CL)\ 
\end{eqnarray}
	on the scale $\Lambda$ of the new physics 
	responsible for the anomalous magnetic moment.

\noindent{\bf Electron}: 
The bound (\ref{mu,d<}) with $\Lambda_{\rm min}$=1TeV implies  
\begin{eqnarray}
	|\delta a_e| < O(1.5\times10^{-10}),\ \ 
	|d_e| < O (3\times10^{-21}e\rm cm), 
\end{eqnarray}
where the former is less stringent than 
	the experimental-theoretical result \cite{mu_e}
\begin{eqnarray}
	\delta a_e = (-1.2\pm2.8)\times10^{-11},    \label{exp_a_e}
\end{eqnarray}
and the latter is much less stringent than the experimental bound \cite{d_e}
\begin{eqnarray}
	d_e = (1.8\pm1.6) \times10^{-27}e\rm cm.    \label{exp_d_e}
\end{eqnarray}
Using the bound (\ref{kL<}) and the 95\%CL upper bounds 
	from (\ref{exp_a_e}) and (\ref{exp_d_e}), we get 
\begin{eqnarray}
	|\mu_e|\Lambda  <O(0.00003), \ \ \ 
	|d_e|\Lambda  <O(6\times10^{-8}), \ \ \ 
\end{eqnarray}
which justifies the pertubation expansions in $\mu_e$ and $d_e$.

\noindent{\bf Tau-lepton}:
The bound (\ref{mu,d<}) with $\Lambda_{\rm min}$=1TeV imply 
\begin{eqnarray}
	|\delta a_\tau| < O(0.002),\ \ 
	|d_\tau| < O (1.0\times10^{-17}e\rm cm), 
\end{eqnarray}
which are more stringent than 
	the experimental results (95\%{\rm CL}) \cite{mud_tau}
\begin{eqnarray}
	&&-0.052<\delta a_\tau<0.058,    \label{exp_a_tau}
\\	&&-3.1\times10^{-16} e {\rm cm}<d_\tau
	<3.1\times10^{-16} e {\rm cm}    \label{exp_d_tau}
\end{eqnarray}
From (\ref{kL<}) with the 95\%CL upper bounds 
	from (\ref{exp_a_tau}) and (\ref{exp_d_tau}), we get 
\begin{eqnarray}
	|\mu_\tau|\Lambda  <O(0.9), \ \ \ 
	|d_\tau|\Lambda  <O(0.9). \ \ \ 
\end{eqnarray}

For the tau-lepton, the experimental bounds for weak dipole moments
	are also available.
It is straightforward to extend our method to the electroweak theory.
We have only to replace $Q$ in the results by 
$(\pm1/4-Q\sin^2\theta)/\cos\theta\sin\theta $ for $Z$ boson,
and by $1/2\sqrt2\sin\theta $ for $W$ boson , where $\theta$
is the Weinberg angle.

\noindent{\bf Neutrinos}:
Because $Q=0$ for neutrinos, the diagram in Fig.\,\ref{f1} are absent,
	and we do not have the relations (\ref{nb2})--(\ref{L<}).
Instead we should evaluate the two-loop diagrams in Fig.\,\ref{f2}.
This may require not only complex calculations, 
but also careful considerations about renormalization
of the severely divergent non-renormalizable diagrams.
We will perform the investigation in other place.
For the present purpose of the order-of-magnitude relations,
	it is sufficient to combine the typical one-loop calculations 
	to guess the result 
\begin{eqnarray}
	\delta m = - 3eg^2c\mu\Lambda^2/64\pi^4,\ 
	\delta m_5 = - 3eg^2cd\Lambda^2/64\pi^4,
\end{eqnarray}
where $c$ is a numerical constant of $O(1)$,
	and $g$ is the gauge coupling constant of SU(2)$\rm_L$.
Then, we obtain the naturalness bound
\begin{eqnarray}
	3eg^2|c\kappa|\Lambda^2/64\pi^4<O(m).\ 
	(\kappa=\mu\ {\rm or}\ d)
\end{eqnarray}
Here we again have three interesting phenomenological applications 
	corresponding to (\ref{mu,d<})--(\ref{L<}).
\\(i) With the nearest new-physics scale, $\Lambda_{\rm min}$=1TeV,
\begin{eqnarray}
	|\mu|,|d|<O\left ({64\pi^4m/3eg^2|c|\Lambda_{\rm min}^2}\right) . 
\label{mu,d_nu<}
\end{eqnarray}
(ii) If we know the experimental upper bound $|\mu|_{\rm max}$ or $|d|_{\rm max}$
	for $|\mu|$ or $|d|$, we have
\begin{eqnarray}
	|\kappa|\Lambda <
	O\left (\sqrt{64\pi^4m|\kappa|_{\rm max}/3e g^2|c|}\right).\
	(\kappa=\mu {\rm\ or\ } d) 
\label{kL_nu<}
\end{eqnarray}
(iii) If we know the experimental lower bound $|\mu|_{\rm min}$ or $|d|_{\rm min}$
	for $|\mu|$ or $|d|$, we have
\begin{eqnarray}
	\Lambda <
	O\left (\sqrt{64\pi^4m/3eg^2|c||\kappa|_{\rm min}}\right).\ \ \ 
	(\kappa=\mu {\rm\ or\ } d) 
\label{L_nu<}
\end{eqnarray}

We use (\ref{mu,d_nu<}) with $\Lambda_{\rm min}$=1TeV
	and the experimental upper bounds \cite{PDG} 
\begin{eqnarray}
	m_{\nu_1}<3.0{\rm eV},\ \  
	m_{\nu_2}<0.19{\rm MeV}, \ \   
	m_{\nu_3}<18.2{\rm MeV} ,
\label{mnu}
\end{eqnarray}
	where $\nu_1$, $\nu_2$ and $\nu_3$, are
	the mass eigenstates of $\nu_e$, $\nu_\mu$ and $\nu_\tau$.
Then we get the naturalness bounds
\begin{eqnarray}
&&	|\mu_{\nu_1}| <O(1.7\times10^{-13}\mu_{\rm B}), \ 
	|d_{\nu_1}| < O(3\times10^{-24}e{\rm cm}), \hskip-10mm 
\cr&&	|\mu_{\nu_2}| < O(1.1\times10^{-8}\mu_{\rm B}), \ 
	|d_{\nu_2}| < O(2\times10^{-19}e{\rm cm}), \hskip-10mm 
\cr&&	|\mu_{\nu_3}| < O(1.1\times10^{-6}\mu_{\rm B}),\ 
	|d_{\nu_3}| < O(2\times10^{-17}e{\rm cm}), \hskip-1mm 
\label{NBmunu}
\end{eqnarray}
which are compared with the experimental or phenomenological bounds
	\cite{mu_nu1} --\cite{d_nu3} 
\begin{eqnarray}
&&	|\mu_{\nu_1}| <1.5\times10^{-10}\mu_{\rm B},\ 
	|\mu_{\nu_2}| <7.4\times10^{-10}\mu_{\rm B},\ 
\cr&&	|\mu_{\nu_3}| <18.2\times10^{-7}\mu_{\rm B},\ 
	|d_{\nu_3}| <5.2\times10^{-17}e{\rm cm},\ 
\label{expnu}
\end{eqnarray}
where the first three are at 90\%CL, and the last, at 95\%CL.
If we use the naturalness bound (\ref{kL_nu<}) 
	and the experimental bounds (\ref{expnu}), we get
\begin{eqnarray}
&&	|\mu_{\nu_1}|\Lambda <O(1.5\times10^{-6}),\ 
	|\mu_{\nu_2}|\Lambda <O(0.0008),\ 
\cr&&	|\mu_{\nu_3}|\Lambda <O(0.4),\ 
	|d_{\nu_3}|\Lambda <O(0.5).\ 
\end{eqnarray}

The experimental results on the solar \cite{solar} 
	and atmospheric \cite{atmo} neutrinos 
	suggest that the differences of $m_{\nu_1}$, $m_{\nu_2}$ and $m_{\nu_3}$
	are much less than the order of eV \cite{mnu}, which implies that 
\begin{eqnarray}
	m_{\nu_1},\ m_{\nu_2}, \ m_{\nu_3}<3.0{\rm eV}.
\label{mnu2}
\end{eqnarray}
If we use (\ref{mnu2}) instead of (\ref{mnu}), we have
\begin{eqnarray}
&&	|\mu_{\nu_1}|,\ |\mu_{\nu_2}|,\ |\mu_{\nu_3}|
	<O(1.7\times10^{-13}\mu_{\rm B}), \ 
\cr&&	|d_{\nu_1}|,\ |d_{\nu_2}|,\ |d_{\nu_3}| 
	< O(3\times10^{-24}e{\rm cm}), 
\end{eqnarray}
instead of (\ref{NBmunu}).

\noindent
{\bf Quarks}:
Though the magnetic and electric dipole moments of quarks 
	are not directly measurable,
	they could affect hadron phenomenology, 
	for example, through scaling violation in deep inelastic 
	lepton-hadron scattering or the electric dipole moments of nucleons.
For quarks, we can again use the bounds (\ref{mu,d<})--(\ref{L<}),
	because they are electrically charged.
The bound (\ref{mu,d<}) with $\Lambda_{\rm min}=1$TeV 
	and the phenomenological values of masses \cite{PDG}
	$m_{\rm u}=(1-5)$MeV,
	$m_{\rm d}=(3-9)$MeV,
	$m_{\rm s}=(75-170)$MeV,
	$m_{\rm c}=(1.15-1.35)$GeV,
	$m_{\rm b}=(4.0-4.4)$GeV, and
	$m_{\rm t}=(174.3\pm5.1)$GeV,
lead to
\begin{eqnarray}
&&	|\mu_{\rm u}| <O(4\times10^{-6}\mu_{\rm N}),\ 
	|d_{\rm u}| <O(4\times10^{-20}e{\rm cm}),\ 
\cr&&	|\mu_{\rm d}| <O(1.5\times10^{-5}\mu_{\rm N}),\ 
	|d_{\rm d}| <O(1.5\times10^{-19}e{\rm cm}),\ \hskip-15mm
\cr&&	|\mu_{\rm s}| <O(0.0003\mu_{\rm N}),\ 
	|d_{\rm s}| <O(3\times10^{-18}e{\rm cm}),\ 
\cr&&	|\mu_{\rm c}| <O(0.0011\mu_{\rm N}),\ 
	|d_{\rm c}| <O(1.1\times10^{-17}e{\rm cm}),\ 
\cr&&	|\mu_{\rm b}| <O(0.007\mu_{\rm N}),\ 
	|d_{\rm b}| <O(7\times10^{-17}e{\rm cm}),\ 
\cr&&	|\mu_{\rm t}| <O(0.14\mu_{\rm N}),\ 
	|d_{\rm t}| <O(1.5\times10^{-15}e{\rm cm}),
\end{eqnarray}
where $\mu_{\rm N}=e/2m_{\rm p}$ with the proton mass $m_{\rm p}$ 
	is the nuclear magneton.

One of the authors (K.~A.) would like to express his gratitude
	to Professor~Takeo~Inami for useful discussions.

\end{document}